# *How Can Brane World Physics Influence Axion Temperature Dependence, Initial Vacuum States, and Permissible Solutions to the Wheeler-De Witt Equation in Early Universe Cosmology?*

A.W. Beckwith
*abeckwith@UH.edu*

## Abstract

We use an explicit Randall-Sundrum brane world effective potential as congruent with conditions needed to form a minimum entropy starting point for an early universe vacuum state. We are investigating whehter the Jeans instability criteria mandating low entropy, low temperature initial pre inflation state configuration can be reconciled with thermal conditions of temperatures at or above ten to the 12 Kelvin, or higher, when cosmic inflation physics takes over. We justify this by pointing to the Ashtekar, Pawlowski, and Singh (2006) article about a prior universe being modeled via their quantum bounce hypothesis which states that this prior universe geometrically can be modeled via a discretized Wheeler-De Witt equation—with it being the collapsing into a quantum bounce point singularity converse of the present day universe expanding from the quantum bounce point delineated in their calculations. The prior universe would provide thermal excitation into the Jeans instability mandated cooled down initial state, with low entropy, leading to extreme graviton production. This necessitates reconciling the lack of a quantum bounce seen in brane world models with the proof of relic graviton production so provided in the simulation so provided. This is also a way of getting around the get around the fact that conventional cosmological CMB is limited by a barrier as of a red shift limit of about z = 1000, i.e. when the universe was about 1000 times smaller and 100,000 times younger than today as to photons, and to come up with a working model of quintessence scalar fields which permits relic generation of dark matter/dark energy.

# I. INTRODUCTION

Our present paper is in response to suggestions by Dr. Wald [1] (2005), Sean Carroll, and Jennifer Chen [2] (2005), and others in the physics department in the University of Chicago about a Jeans instability criteria leading to low entropy states of the universe at the onset of conditions before inflationary physics initiated expansion of inflaton fields. We agree with their conclusions and think it ties in nicely with the argument so presented as to a burst of relic gravitons being produced. This also is consistent with an answer as to the supposition for the formation of a unique class of initial vacuum states, answering a question Guth [3] raised in 2003 about if or not a preferred form of vacuum state for early universe nucleation was obtainable. This is in tandem with the addition of gravity changing typical criteria for astrophysical applications of the jeans instability criteria [4] for weakly interacting fields, as mentioned by Penrose [49].

Contemporary graviton theory states as a given that there is a thermal upsurge which initiates the growth of graviton physics. This is shown in K.E. Kuntz's well written (2002) article [5, which gives an extremely lucid introduction as to early universe additional dimensions giving a decisive impetus to giving additional momentum to the production of relic gravitons. However, Kunze [5] is relying upon enhanced thermal excitation states, which contradict the Jeans instability criteria which appears to rule out a gravitational field soaked initial universe configuration being thermally excited. Is there a way to get around this situation which appears to violate the Jeans instability criteria for gravitational fields/gravitons in the early universe mandating low entropy states? We believe that there is—and that it relies upon a suggestion given by Ashtekar, Pawlowski, and Singh [6,7] (2006) regarding the influence of the quantum bounce via quantum loop gravity mirror imaging a prior universe collapsing into a 'singularity' with much the same geometry as the present universe. If this is the case, then we suggest that an energy flux from that prior universe collapse is transferred into a low entropy thermally cooled down initial state, leading to a sudden burst of relic gravitons as to our present universe configuration. The first order estimate for this graviton burst comes from the numerical density equation for gravitons written up by Weinberg as of 1971 [8] with an exponential factor containing a frequency value divided by a thermal value, T, minus 1. If the frequency value is initially quite high, and the input given by a prior universe 'bounce', with an initial very high value of energy configuration, then we reason that this would be enough to introduce a massive energy excitation into a thermally cooled down axion wall configuration which would then lead to the extreme temperatures of approximately $10^{12}$ Kelvin forming at or before a Planck interval of time $t_P$, plus a melt down of the axion domain wall, which we then says presages formations of a Guth style inflationary quadratic and the onset of chaotic inflationary expansion.

A way of getting to all of this is to work with a variant of the Holographic principle, and an upper bound to entropy calculations. Busso and Randall [9] (2001) give a brane world variant of the more standard upper bounds for entropy in terms of area calculations times powers of either the fourth or fifth dimensional values of Planck mass [45] (45), which still lead to minimized values if we go near the origins of the big bang itself. Our observations are then not only consistent with the upper bound shrinking due to smaller and smaller volume/area values of regions of space containing entropy measured quantities, but consistent with entropy/area being less than or equal to a constant times absolute temperatures, if we take as a given in the beginning low temperature conditions prior to the pop up of an inflation scalar field.

Recently, Feng et. al [10], introduced the idea of an effective Lagrangian to compliment the idea of CPT violations, as new physics, composed of a term proportional to the derivative of a scalar field $\phi$ (in this case a quintessence field) times the dual of the electromagnetic tensor. What we are supplying is a proof if you will of time dependence of the quintessence scalar field, and Feng et al 10] inputs into the electric and magnetic fields of this dual of the E&M tensor from the stand point of CMB. It is noteworthy to bring up that Ichiki et al [11] notes that because standard electromagnetic fields are conformally coupled to gravity, magnetic fields simply dilute away as the universe expands, i.e. we need to consider the role of gravity generation in early universe models. We will, in this document try to address how, via graviton production, we have intense gravity wave generation, and also how to use this as a probe of early universe quintessence fields, and also how to get around the fact that conventional cosmological CMB is limited by a barrier as of a red shift limit of about $z \approx 1000$, i.e. when the universe was about 1000 times smaller and 100,000 times younger than today as to photons ,i.e. we are confirming as was stated by Weinberg as of 1977 that there is zero chance of relic photon generation from the big bang itself we can see being observable is zero and that we are using relic gravitons as a probe as to the physics of quintessence fields , as well as the origins of dark matter/dark energy issue.

In addition this approach accounts for data suggesting that the four-dimensional version of the "cosmological constant" in fact varies with respect to external background temperature. If this temperature significantly varied during early universe baryogenesis, the end result is that there would be a huge release of spin-two gravitons in the early stages of cosmic nucleation of a new universe. It also answers whether "Even if there are $10^{1000}$ vacuum states produced by String theory, then does inflation produces overwhelmingly one preferred type of vacuum states over the other possible types of vacuum states?" [12] (Guth, 2003).

Finally, but not least, we also account for the evolution of an equation of motion of a quintessence field, via equations given to us by M. Li, X. Wang, B. Feng, and X. Zhang [13] , which is a first ever re do in dept of the interaction of a quintessence scalar field with baryonic 'normal matter' assuming varying contributions to a potential field system with a varying by temperature axion mass contribution to an evolving pre inflationary state, which collapses to a quadratic Guth style inflationary state with a suitable rise in initial inflationary temperatures.

## II. ORGANIZATION OF THIS PAPER

## III. REVIEW OF WHAT CAN BE IDENTIFIED VIA SCALING ARGUMENTS AND VARYING MODELS OF DARK ENERGY

We will now review the scaling arguments as to permissible entropy behavior and use this to begin our inquiry as to what to expect from brane models as given by Alam and Strobinsky et al in July 2004 [14]. To begin, they summarize several dark energy candidates as having the following inquiry schools, which I will reproduce in part for different equations of state, $w(z) = \frac{P}{\rho}$ as the ratio of pressure over observed density of states

(1) Dark energy with $w(z) \leq -1$

(2) Chaplygin Gas models with $w(z) = 0$ for high red shift to $w(z) \cong -1$ today

(3) Brane world models where acceleration, cosmology wise, is due to the gravity sector, rather than matter sector

(4) Dark energy models with negative scalar potentials

(5) Interacting models of dark energy and dark matter

(6) Modified gravity and scalar-tensor models

(7) Dark energy driven by quantum effects

(8) Dark energy with a late time transition in the equation of state

(9) Unified models of dark energy and inflation

The model they ultimately back in part due to astro physics observations is closest to one with $w(z) = 0$ in the distant past, to one with $w(z) \cong -1$ today. We will next go to scaling argument in part to talk about the significance of such thinking in terms of entropy. The model results they have in initial cosmology is not significantly different in part from the modeled values obtained by Knop et al with $-1.61 < w(z) < -.78$ and in some particulars are close to what the Chaplygin Gas model predicts when dark energy—dark matter unification is achieved through an exotic background fluid whose equation of state is given by p = - A/$\rho^\alpha$ , and with $0 < \alpha \leq 1$ We are not specifically endorsing this model, but are using the equation of state values to investigate some fundamental initial conditions for vacuum nucleation and brane world cosmology. We should note that if we consider $w(z) \cong -1$ we are introducing driven inflation via cosmological constant models.

We begin with scale models, which we claim break down in part as follows:

With

$$\frac{\ddot{a}}{a} \approx -\frac{4}{3} \cdot \rho \cdot \left[1 + \frac{3}{w(z)}\right] \tag{1}$$

The generalized Chaplygin gas (GCG) model allows for a unified description of the recent accelerated expansion of the Universe and the evolution of energy density perturbations. If we use $w(z) \propto \varepsilon^+$, we have the following,

$$\frac{\ddot{a}}{a} \approx -\frac{4}{3} \cdot \rho \cdot \left[\left(1 + \frac{3}{\varepsilon^+}\right) \equiv N^+\right] \tag{2}$$

If we have a situation for which

$$\rho \equiv \rho_0 \cdot e^{-C \cdot t_1} \tag{3}$$

Before proceeding on applying the third equation, we need to show how it ties in with the Chaplygin Gas model predictions, and generalized fluid models. Begin first with a density varying as, due to a red shift $z \equiv 1/(a-1)$

$$\rho_i = \rho_{i0} \cdot (1+z)^{3 \cdot (1+w_i(z))} \qquad (4)$$

This is in tandem with the use of, for $H = \dot{a}/a$ and an $i^{th}$ density parameter of $\Omega_i \equiv \rho_{i0}/\rho_C$ and $\rho_C$ a critical density parameter, with $\Omega_i \equiv \rho_{i0}/\rho_C \xrightarrow[w \to -1]{} \Lambda$ (like a cosmological constant)

$$H^2 \equiv \sum_i \Omega_i \cdot a^{-3 \cdot (1+w_i)} + (1 - \sum_i \Omega_i) \cdot a^{-2} \qquad (5)$$

The upshot is that if $w_i \approx 0 \Leftrightarrow .8 \leq z \leq 1.75$ which occurs if time $t$ is picked for $t_{present} \gg t \geq t_P$

$$\rho \sim \rho_{i0} \cdot (1+z)^2 \leq 8 \cdot \rho_{i0} \qquad (6)$$

Versus the later time estimate of, close to the present era of z =0 and w almost = -1

$$\rho \sim \rho_{i0} \qquad (7)$$

I.e. there was a major drop off of density values from earlier conditions to the present era. And this is not even getting close to the density values one would have for times $t \leq t_P$ which we will comment upon later. Given this though, let us now look at some consequences of this drop off of density

We can consider a force on the present 'fluid' constituents of a joint dark energy-dark mass model.

$$-\frac{dV}{da} = -(1 + 3 \cdot w_i) \cdot \Omega_i \cdot a^{-(2+3 \cdot w_i)} \qquad (8)$$

When we have a small interval of time after $t \geq t_P$, we have $w_i \approx \varepsilon^+$ leading to, for small values of the scale factor $a(t) \sim \delta^+$, and a potential system we call $V_{early}$ for early universe scalar field conditions

$$\frac{dV_{early}}{da} = (1 + 3 \cdot \varepsilon^+) \cdot \Omega_i \cdot (a \sim \delta^+)^{-2} \approx \Omega_i / \delta^+ \qquad (9)$$

This implies a large force upon any structure in the early universe which so happens to be accurate.

We can contrast this with, for $w_i \approx -1$

$$\frac{dV_{Today}}{da} = -(2) \cdot \Omega_i \cdot (a \sim Big)^{-2} \approx -N^{++} \Omega_i \qquad (10)$$

This is implying a large positive force leading to accelerated expansion, whereas Eqn. (9) predicts at or before time $t_P$ a negative force which would be consistent with early universe pre big bang conditions. Furthermore, we can also look at what this implies for the Friedman equations with respect to the scale factor at (or before) Planck's time $t_P$, i.e.

$$\frac{\ddot{a}}{a} \approx -\frac{4}{3} \cdot \rho_0 \cdot [\exp(-C_1 t)] \cdot [N^+] \tag{11}$$

If we for small time intervals look at $a \sim t^{1/3+\tilde{\gamma}}$, then Eqn. (11) above reduces to for times near the Planck unit

$$-(2/9) \cdot t_p^{-2} \approx -4 \cdot G \cdot \rho_0 \cdot N^+ / \left(3 \cdot \left[ (1 + C_1 \cdot t_P) = \varepsilon^+ \right] + C_1^2 \cdot t_P^2 / 2 \right]) \tag{12}$$

This leads to

$$C_1 = \frac{-1 + \varepsilon^+}{t^P} \tag{13}$$

This is in the neighborhood of Plank unit time confirmation of the graceful exit from inflation, i.e. a radical negative acceleration value we can write as

$$\frac{\ddot{a}}{a} \approx -\frac{4 \cdot G \cdot \rho_0}{3} \cdot N^+ \tag{14}$$

As well as a provisional density behavior we can write as

$$\rho \sim \rho_0 \cdot \exp\left[ \frac{-\varepsilon^+(small \quad time) + t}{t_P} \right] \tag{15}$$

If we have a situation for which time is smaller than the Planck interval time, we have Eqn. (15) predicting that there is decreasing density values, and that Eqn. (15) would predict peak density values at times $t \approx t_P$, which in a crude sense is qualitatively similar to the picture we will outline later of a nucleation of a vacuum state leading to a final nucleated density. This however, also outlines the limits of the Friedman equation for early universe cosmology. It is useful to note though that should one pick $w_i \geq -1$ as is indicated is feasible in the observational sense that Eqn. 1 above predicts a positive right hand side implying positive acceleration of scale factors. This is akin to the gas model predicting increased acceleration in the present cosmological era.

We now should now look at the role entropy plays in early universe nucleation models. This will in its own way be akin to making sense of the discontinuity in cosmological scale factors for expansion seen in our future axion wall model where we will write varying scalar model potentials which will indicate in Eqn. (25) to Eqn. (27) below (47)

$$V \approx (1/n) \cdot \phi^n \xrightarrow[AXION \to 0]{} (1/2) \cdot \phi^2 \tag{16}$$

This would be in tandem with

$$a_{init} \cdot \exp\left( \frac{4 \cdot \pi}{n} \cdot \left( \phi^2_{init} - \phi^2(t) \right) \right) \xrightarrow[AXION \to 0]{} a_{init} \cdot \exp\left( \frac{4 \cdot \pi}{2} \cdot \left( \phi^2_{init} - \phi^2(t) \right) \right) \tag{17}$$

## IV. RE CONSTRUCTING WHAT CAN BE SAID ABOUT INITIAL VACUUM FLUCTUATIONS AND THEIR LINKAGE TO BRANE WORLD PHYSICS

We shall reference a simple Lypunov Exponent argument as to adjustment of the initial quantum flux on the brane world picture. This will next be followed up by a description of how to link the estimated requirement of heat influx needed to get the quantum spatial variation flux in line with inflation expansion parameters.

To begin this, we access the article "Quantum theory without Measurements [15] "to ascertain the role of a Lyapunov exponent $\widetilde{\Lambda}_{chaos}$ such that

$$\Delta p = (\Delta p_0) \cdot \exp(-\widetilde{\Lambda}_{chaos} \cdot t) \tag{18}$$

And

$$\Delta x = (\hbar/\Delta p_0) \cdot \exp(\widetilde{\Lambda}_{chaos} \cdot t) \tag{19}$$

Here, we define where a wave functional forms via the minimum time requirement as to the formation of a wave functional via a minimum time of the order of Planck's time

$$t_h = (\widetilde{\Lambda}_{chaos})^{-1} \cdot \ln[\Delta p_0 \cdot \widehat{L}/\hbar] \approx t_P \tag{20}$$

If we have a specified minimum length as to how to define $\widehat{L} \approx l_P$, this is a good way to get an extremely large $\widetilde{\Lambda}_{chaos}$ value, all in all so that we have Eqn. 19 above on the order of magnitude at the end of inflation as large as what the universe becomes, i.e. a few centimeters or so, from an initial length value on the order of Planck's length $\widehat{L} \approx l_P$.

## V. WHY ADDRESS A SIMPLIFIED FLUCTUATION PROCEDURE

Two reasons  First of all, we have that  our description of a link of the sort between a brane world effective potential and Guth style inflation has been partly replicated by Sago, Himenoto, and Sasaki in November 2001[16] where they assumed a given scalar potential, assuming that m is the mass of the bulk scalar field

$$V(\phi) = V_0 + \frac{1}{2}m\phi^2 \quad \dots\dots\dots\dots\dots\dots\dots\dots\dots\dots\dots\dots\dots\dots\dots\dots\dots\dots\dots\dots\dots. \tag{21}$$

Their model is in part governed by a restriction of their 5-dimensional metric to be of the form, with $l =$ brane world curvature radius, and H their version of the Hubble parameter

$$dS^2 = dr^2 + (H \cdot l)^2 \cdot dS^2_{4-\dim} \tag{22}$$

I.e. if we take $k_5^2$ as being a 5 dimensional gravitational constant

$$H = \frac{k_5^2 \cdot V_0}{6} \quad \dots\dots\dots\dots\dots\dots\dots\dots\dots\dots\dots\dots\dots\dots\dots\dots\dots\dots\dots\dots. \tag{23}$$

Our difference with Eqn. (22) is that we are proposing that it is an intermediate step, and not a global picture of the inflation field potential system. However, the paper they present with its focus upon the zero mode contributions to vacuum expectations $\langle \delta\phi^2 \rangle$ on a brane has similarities as to what we did which should be investigated further. The difference between what they did, and our approach is in their value of

$$dS^2_{4-\dim} \equiv -dt^2 + \frac{1}{H^2} \cdot [\exp(2 \cdot H \cdot t)] \cdot dx^2 \qquad (24)$$

Which assumes one is still working with a modified Gaussian potential all the way through, as seen in Eqn. (21). This is assuming that there exists an effective five dimensional cosmological parameter which is still less than zero, with $\Lambda_5 < 0$, and $|\Lambda_5| > k_5^2 \cdot V_0$ so that

$$\Lambda_{5,eff} = \Lambda_5 + k_5^2 \cdot V_0 < 0 \qquad \ldots \ldots (25)$$

It is simply a matter of having

$$|m^2| \cdot \phi^2 << V_0 \qquad \ldots \ldots (26)$$

And of making the following identification

$$\phi_{5-\dim} \propto \tilde{\tilde{\phi}}_{4-\dim} \equiv \tilde{\tilde{\phi}} \approx [\phi - \varphi_{fluctuations}]_{4-\dim} \qquad \ldots \ldots (27)$$

With $\varphi_{fluctuations}$ in Eqn. (27) is an equilibrium value of a true vacuum minimum for a chaotic four dimensional quadratic scalar potential for inflationary cosmology. This in the context of the fluctuations having an upper bound of $\tilde{\tilde{\phi}}$ (Here, $\tilde{\tilde{\phi}} \geq \varphi_{fluctuations}$). And $\tilde{\tilde{\phi}}_{4-\dim} \equiv \tilde{\tilde{\phi}} - \frac{m}{\sqrt{12 \cdot \pi \cdot G}} \cdot t$, where we use $\tilde{\tilde{\phi}} > \sqrt{\frac{60}{2 \cdot \pi}} M_P \approx 3.1 M_P \equiv 3.1$, with $M_P$ being a Planck mass. This identifies an imbedding structure we will elaborate upon later on. However, in doing this, Eqn. (21) is ignoring axion walls which make the following contribution to cosmology (where does dark matter-dark energy come from?) I.e. we look at axion walls specified by Kolb's book [17] conditions in the early universe (1991) with his Eqn. (10.27) vanishing and collapsing to Guth's quadratic inflation. I.e. having the quadratic contribution to an inflation potential arise due to the vanishing of the axion contribution of the first potential of Eqn. (8) above with a temperature dependence of

$$V(a) = m_a^2 \cdot (f_{PQ}/N)^2 \cdot (1 - \cos[a/(f_{PQ}/N)]) \qquad \ldots \ldots (28)$$

Here, he has the mass of the axion potential as given by $m_a$ as well as a discussion of symmetry breaking which occurs with a temperature $T \approx f_{PQ}$. This is done via scaling the axion mass via either [18]

$$m_a(T) \approx 3 \cdot H(T) \qquad (29)$$

So that the axion 'matter' will oscillate with a 'frequency' proportional to $m_a(T)$. The hypothesis so presented is that input thermal energy given by the prior universe being inputted into an initial cavity / region dominated by an initially configured low temperature axion domain wall would be thermally excited to reach the regime of temperature excitation permitting an order of magnitude drop of axion density $\rho_a$ from an initial temperature $T_{dS}|_{t \leq t_P} \sim H_0 \approx 10^{-33} eV$ as given by( assuming we will use the following symbol $\psi_a$ for axions, and then relate it to Guth inflationary potential scalar fields later on, and state that $\psi_a(t) = \psi_i$ is the initial misaligned value of the field)

$$\rho_a(T_{ds}) \propto \frac{1}{2} \cdot m_a(T_{dS}) \cdot \psi_i^2 \xrightarrow[T \to 10 \text{ to } 12th \text{ power Kelvin}]{} \varepsilon^+ \qquad (30)$$

Or

$$m_a(T) \cong 0.1 \cdot m_a(T=0) \cdot (\Lambda_{QCD}/T)^{3.7} \quad \text{.........................................} \quad (31)$$

The dissolving of axion walls is necessary for dark matter-dark energy production and we need to incorporate this in a potential system in four dimensions, and relate it to a bigger five dimensional potential systems. First of all though we need to find a way to, using brane theory, to investigate how we can have non zero axion mass conditions to begin with. Needless to say though, to explain what we are doing via quantum fluctuations, we need to use them to obtain a working description of how to link thermal inputs into our potential system, in line with what we will develop later on in this manuscript. Let us now talk about how to input thermal heat up of the axion walls

## VI. THERMAL HEAT UP OF THE AXION WALLS JUST PRESENTED VIA QUANTUM FLUCTUATIONS

Recalling our equation for $t_h$, i.e. the time when a wave function could form, we have if we look at (37) $\widehat{L} \approx l_P \equiv$ Planck's length, a criteria for getting a VERY large value for the Lypunov coefficient $\widetilde{\Lambda}_{chaos}$ which we used to justify an inflation congruent set of values for $\Delta x$. It is now time to consider what thermal flux input could be expected to make $\Delta x$ expand so fast up to $10^{-23}$ seconds to be on the order of a grapefruit in size. To do this, we need to look at the zero energy density $\rho_\Lambda$ which comes from

$$(4 \cdot \pi/3) \cdot \widehat{L}^3 \cdot \rho_\Lambda \propto 4 \cdot \pi \cdot \widehat{L} \cdot M_P^2 \quad \text{..................................} \quad (32)$$

So that we can write as a criteria for <u>when</u> the wave functional will actually form, i.e. [100]

$$t_h \approx \widetilde{\Lambda}_{chaos}^{-1} \cdot \ln\left[\frac{\Delta p_0}{\hbar} \cdot \frac{c_1}{3} \cdot \left[Energy \quad input\right]\right] \quad \text{...............................} \quad (33)$$

This presumes we have a criterion for input of energy into determining when a wave functional would form. The existence of a wave functional in this situation would lend credence to the LQG work with revised versions of the Wheeler De Witt equation. Needless to say though we have to consider that we do not have a criterion written explicitly yet as to how to get a low temperature initial set of conditions for the inclusion of this energy input. We shall next provide such criteria based upon Brane world physics.

## VII. SETTING UP CONDITIONS FOR ENTROPY BOUNDS VIA BRANE WORLD PHYSICS

Our starting point here is first showing equivalence of entropy formulations in both the Brane world and the more typical four dimensional systems. A Randall-Sundrum Brane world will have the following as a line element and we will continue from here to discuss how it relates to holographic upper bounds to both anti De sitter metric entropy expressions and the physics of dark energy generating systems.

To begin with, let us first start with the following as a $A \cdot dS_5$ model of tension on brane systems, and the line elements. If there exists a tension $\widecheck{T}$, with Plank mass in five dimensions denoted as $M_5$, and a curvature value of $l$ on $A \cdot dS_5$ we can write [19]

$$\widecheck{T} = 3 \cdot \left(M_5^3/4 \cdot \pi \cdot l\right) \quad (34)$$

Furthermore, the $A \cdot dS_5$ line element, with $r =$ distance from the brane, becomes [14]

$$\frac{dS^2}{l^2} = (\exp(2 \cdot r)) \cdot \left[-dt^2 + d\rho^2 + \sin^2 \rho \cdot d\Omega_2\right] + dr^2 \quad \dots \dots \dots \dots \dots \dots \dots \dots \quad (35)$$

We can then speak of a four dimensional volume $V_4$ and its relationship with a three dimensional volume $V_3$ via

$$V_4 = l \cdot V_3 \quad (36)$$

And if a Brane world gravitational constant expression $G_N = M_4^{-2} \Leftrightarrow M_4^2 = M_5^3 \cdot l$ we can get a the following space bound Holographic upper bound to entropy

$$S_5(V_4) \leq V_3 \cdot (M_5^3 / 4) \quad (37)$$

If we look at an area 'boundary' $A_2$ for a three dimensional volume $V_3$, we can re cast the above holographic principle to (for a volume $V_3$ in Planck units)

$$S_4(V_3) \leq A_2 \cdot (M_4^2 / 4) \quad (38)$$

We link this to the principle of the Jeans inequality for gravitational physics and a bound to entropy and early universe conditions, as given by S. Carroll and J.Chen (2005) [15] stating if $S_4(V) = S_5(V_4)$ then if we can have

$$A_2 \xrightarrow[t \to t_P]{} \varepsilon_{small\ area} \Leftrightarrow S_5(V_4) \approx \delta_{small\ entropy} \quad (39)$$

Low entropy conditions for initial conditions, as stated above give a clue as to the likely hood of low temperatures as a starting point via R. Easther et al. (1998) [20] assuming a relationship of a generalized non brane world entropy bound, assuming that $n^* \approx$ bosonic degrees of freedom and $T$ as generalized temperature, so we have as a temperature based elaboration of the original work by Susskind [21] holographic projections forming area bound values to entry

$$\frac{S}{A} \leq \sqrt{n^*} \cdot T \quad (40)$$

Similar reasoning, albeit from the stand point of the Jeans inequality and instability criteria lead to Sean Carroll and J. Chen (2005) [2] having for times at or earlier than the Planck time $t_P$ that a vacuum state would initially start off with a very low temperature

$$T_{ds}\big|_{t \leq t_P} \sim H_0 \approx 10^{-33} eV \quad (41)$$

We shall next refer to how this relates to, considering a low entropy system an expression Wheeler wrote for graviton production and its implications for early relic graviton production, and its connection to axion walls and how they subsequently vanish at or slightly past the Planck time $t_P$.

This will in its own way lead us to make sense of a phase transition we will write as a four dimensional embedded structure within the Sundrum brane world structure

$$\begin{aligned}\tilde{V}_1 &\to \tilde{V}_2 \\ \tilde{\phi}(increase) \leq 2\cdot\pi &\to \tilde{\phi}(decrease) \leq 2\cdot\pi \\ t \leq t_P &\to t \geq t_P + \delta\cdot t\end{aligned} \quad (42)$$

The potentials $\tilde{V}_1$, and $\tilde{V}_2$ will be described in terms of **S-S'** di quark pairs nucleating and then contributing to a chaotic inflationary scalar potential system. Here, $m^2 \approx (1/100)\cdot M_P^2$

$$\tilde{V}_1(\phi) = \frac{M_P^2}{2}\cdot(1-\cos(\tilde{\phi})) + \frac{m^2}{2}\cdot(\tilde{\phi}-\phi^*)^2 \quad (43)$$

$$\tilde{V}_2(\phi) \propto \frac{1}{2}\cdot(\tilde{\phi}-\phi_C)^2 \quad (44)$$

## VIII. LINKS TO DE COHERENCE AND HOW WE GO FROM EQN. 43 TO EQN. 44 ABOVE

Recall what was said earlier as to the relationship between $\rho_a(T_{dS}) \propto \frac{1}{2}\cdot m_a(T_{dS})\cdot \psi_i^2 \xrightarrow[T\to 10\ to\ 12th\ power\ Kelvin]{} \varepsilon^+$ and the formation of Guth style inflation as represented by Eqn. (44) above. Here we first have to consider when the effects of thermal input into the geometry given by Eqn. 43 above are purely quantum mechanical. That was given by Weinberg [8] (1977) via

$$\alpha_g \approx \left[\frac{G\cdot E_{input}^2}{\hbar}\right] \propto O(1) \equiv \text{Order of unity} \quad (45)$$

This requires an extremely high thermal input temperature of the order of magnitude of nearly $10^{+34}\,Kelvin$, far higher than the nucleosynthesis values of $10^{+12}\,Kelvin$, so we will discuss how to put in measures of coherence and de coherence between a scalar field value, and the mass of an axion wall to discuss how $\rho_a(T_{dS}) \xrightarrow[T\to 10\ to\ 12th\ power\ Kelvin]{} \varepsilon^+$

I.e. we look at de coherence specified via

$$\langle e^{i\phi}\rangle \approx \exp[-\gamma_{De-coherence}\cdot t] \quad \dots\dots (46)$$

Where we write

$$\gamma_{De-Coherence} \cong \frac{\eta\cdot(\hat{L}\approx l_P)^2\cdot T_{input\ temperature}}{\hbar} \quad \dots\dots (47)$$

The mass of an axion wall is tied in with the vibrational frequency, as noted in the case where one supposes that the axion 'matter' will oscillate with a 'frequency' proportional to $m_a(T)$. If so then if we look at a time of the order of Planck's time $t_P$. I.e. if we consider an initial current via an ohmic 'current' $J$ for early universe flux of materials from an initial nucleation point

$$J \approx \eta \cdot \omega_{eff} \propto \eta \cdot \frac{E_{eff}}{\hbar} \approx \eta \cdot \frac{E_{Thermal-Energy}}{\hbar} \tag{48}$$

We get

$$\exp[-\gamma_{De-coherence} \cdot t_P] \approx 0 \quad \text{If} \quad T_{thermal-input} > 10^{12} \, Kelvin \tag{49}$$

$$\exp[-\gamma_{De-coherence} \cdot t_P] \approx 1 \quad \text{If} \quad T_{thermal-input} << 10^{12} \, Kelvin \tag{50}$$

Having said this, we now can consider a thermal by product, i.e. intense gravitation production as a side product of this change in de coherence values. This will require, though having low temperature, low entropy values to start with, and we will examine this after we discuss how gravitons could be produced

## IX. THE WHEELER GRAVITON PRODUCTION FORMULA FOR RELIC GRAVITONS

As is well known, a good statement about the number of gravitons per unit volume with frequencies between $\omega$ and $\omega + d\omega$ may be given by (assuming here, that $\bar{k} = 1.38 \times 10^{-16} \, erg/^0 K$, and $^0 K$ is denoting Kelvin temperatures, while we keep in mind that Gravitons have two independent polarization states)

$$n(\omega)d\omega = \frac{\omega^2 d\omega}{\pi^2} \cdot \left[\exp\left(\frac{2 \cdot \pi \cdot \hbar \cdot \omega}{\bar{k} \cdot T}\right) - 1\right]^{-1} \tag{51}$$

This formula predicts what was suggested earlier. A surge of gravitons commences due to a rapid change of temperature. I.e. if the original temperature were low, and then the temperature rapidly would heat up? Here is how we can build up a scenario for just that. Eqn. (51) suggests that at low temperatures we have large busts of gravitons.

Now, how do we get a way to get the $\omega$ and $\omega + d\omega$ frequency range for gravitons, especially if they are relic gravitons? First of all, we need to consider that certain researchers claim that gravitons are not necessarily massless, and in fact the Friedman equation acquires an extra dark-energy component leading to accelerated expansion. The mass of the graviton allegedly can be as large as ~ (1015 cm)-1. This is though if we connect massive gravitons with dark matter candidates, and not necessarily with relic gravitons. Having said this we can note that Massimo Giovannini [22] writes an introduction to his Phys Rev D article about presenting a model which leads to post-inflationary phases whose effective equation of state is stiffer than radiation. He states : The expected gravitational wave logarithmic energy spectra are tilted towards high frequencies and characterized by two parameters: the inflationary curvature scale at which the transition to the stiff phase occurs and the number of (nonconformally coupled) scalar degrees of freedom whose decay into fermions triggers the onset of a gravitational reheating of the Universe. Depending upon the parameters of the model and upon the different inflationary dynamics (prior to the onset of the stiff evolution), the relic gravitons energy density can be much more sizable than in standard inflationary models, for frequencies larger than 1 Hz. Giovannini [22] claims that there are grounds for an energy density of relic gravitons in critical units (i.e., h02ΩGW) is of the order of 10-6, roughly eight orders of magnitude larger than in ordinary inflationary models. That roughly corresponds with what could be expected in our brane world model for relic graviton production.

We also are as stated earlier , stating that the energy input into the frequency range so delineated comes from a prior universe collapse , as modeled by Ashtekar, A., Pawlowski, T. and Singh, P [6,7](2006) via their quantum bounce model as given by quantum loop gravity calculations. We will state more about this later in this document.

Another take on Eqn. (44) is that the domain walls are removed via a topological collapse of domain walls as alluded to by the Bogomolnyi inequality. This would pre suppose that early universe conditions are in tandem with Zhitinisky's (2002) [23] supposition of color super conductors. Those wishing to see a low dimensional condensed matter discussion of applications of such methodology can read my articles in World press scientific, as well as consider how we can form a tunneling Hamiltonian treatment of current calculations. Either interpretation will in its own way satisfy the requirements of baryogenesis, and also give a template as to the formation of dark energy.

I.e. look at conditions for how Eqn. (44) may be linked to a false vacuum nucleation. The diagram for such an event is given below, with a tilted washboard potential formed via considering the axion walls with a small term added on, which is congruent with, after axion wall disappearance with Guth's chaotic inflation model.

The hypothesis so presented is that input thermal energy given by the prior universe being inputted into an initial cavity / region dominated by an initially configured low temperature axion domain wall would be thermally excited to reach the regime of temperature excitation permitting an order of magnitude drop of axion density $\rho_a$ from an initial temperature2 $T_{dS}|_{t \leq t_P} \sim H_0 \approx 10^{-33} eV$

As referred to in V.Mukhanov's book [24] on foundations of cosmology, spalerons are a way to introduce motion of a 'quasi particle' in a Euclidian metric via use of Wick rotations $\tau = -it$ Mukhanov introduces two ways for an instanton (spaleron) to have an escape velocity from a rotated Euclidian metric defined potential, in terms of a given thermal bath of temperature $T$. The two limiting cases are in part defined by the formation of an instanton action $S_I$, with [14]

$$T << V(q_m)/S_I \Rightarrow \text{Rate of escape determined by the instanton} \qquad (52)$$

This assumes

$$E << V(q) \qquad (53)$$

Next

$$T >> V(q_m)/S_I \Rightarrow \text{Period of oscillation is about zero} \qquad (54)$$

This assumes

$$E \approx V(q) \qquad (55)$$

When we have the energy of the system close to Eqn. (55), we are in the realm of a first order approximation of escape probability of constituents of a scalar field $\phi$ given by

$$P \propto \exp(-\frac{E_{sphalerons}}{T}) \qquad (56)$$

## X. INITIAL TEMPLATE FOR POSSIBLE INTERPRETATION OF FORMATION/ DISAPPEARANCE OF AXION WALLS

This assumes that the energy (mass) of a sphaleron [14] is defined via

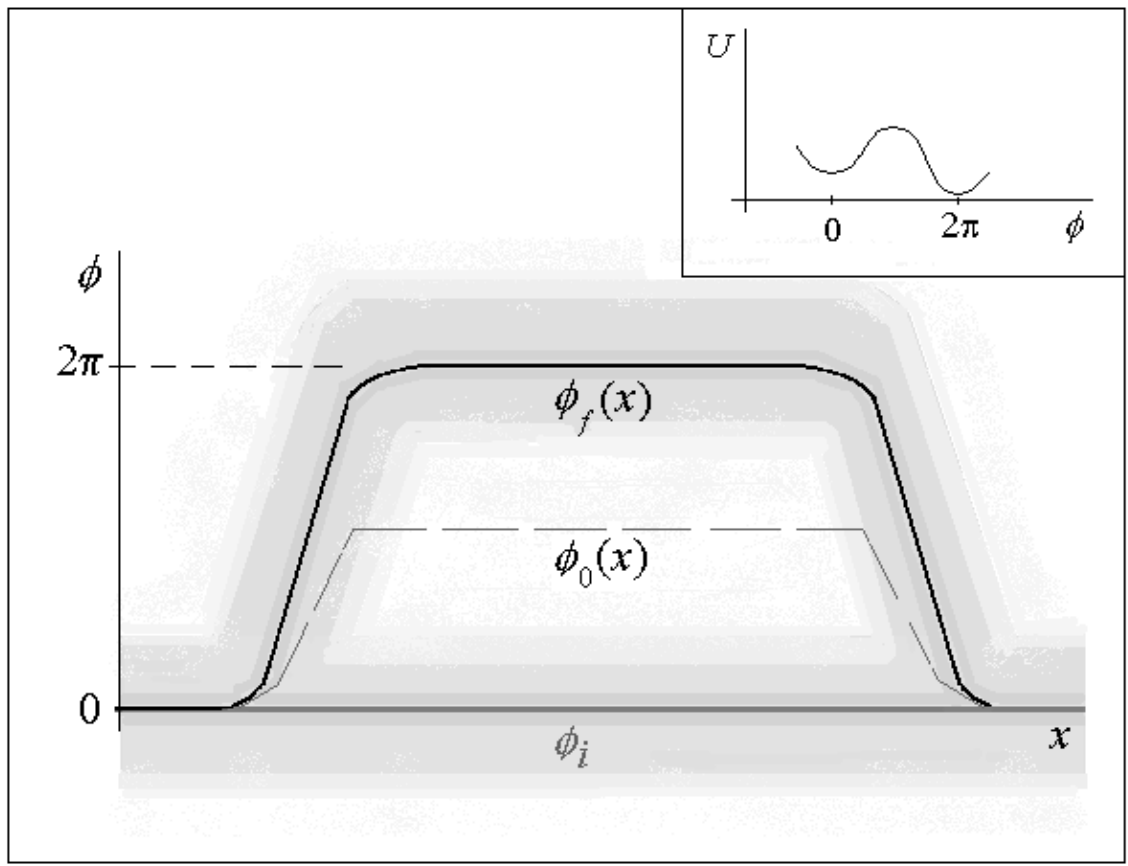

**FIGURE 1**

*Initial set up for nucleation of a Coleman-De Luccia Instanton? Assuming that the initial state so referred to is in tandem with 1) creation of relic gravitons in an initially low temperature environment, and 2) the existence of temperature dependent axion walls. This would either confirm the prediction that Eqn. (30) contributes to a Coleman-De Luccia Instanton if indeed temperatures are initially quite low before a Planckian time interval $t_P$, or give credence to the use of the topological domain wall formation/ subsequent collapse due to the Bogomolnyi inequality*

$$E_{sphalerons} \approx V(q_m) \tag{57}$$

The exact particulars of forming an appropriate instanton $S_I$ for a di quark condensate are yet to be satisfactorily determined, but we will categorically state that the model we are working with assumes a drastic heat up of early nucleation initial conditions to permit after some period of time after a Planck's time $t_P$ conditions permitting chaotic inflation. This after the main contribution to the chaotic inflationary potential 'tilt' turns from an initial mass contribution to energy, allowing for Guth's quadratic potential for scalar fields to be a primary contribution to cosmological initial conditions we can measure

Given this first figure, let us consider a four dimensional potential system, which is for initial low temperatures, and then next consider how higher temperatures may form, and lead to the disappearance of axion walls. To do this we refer to what is written in Eqns. (42) to (44). Note that potentials $\widetilde{V}_1$, and $\widetilde{V}_2$ are two cosmological inflation potential, and $t_P =$ Planck time (the time it would take a photon traveling at the speed of light to cross a distance equal to the Planck length $\approx 5.39121(40) \times 10^{-44}$ seconds; Planck length denoted by $l_P$ is the unit of length approximately $1.6 \times 10^{-35}$ meters; it is in the system of units known as Planck units; the Planck length is deemed "natural" because it can be defined from three fundamental physical constants: the speed of light, Planck's constant, and the gravitational constant).

We are showing the existence of a phase transition between the first and second potentials, with a rising and falling value of the magnitude of the four dimensional scalar fields. When the scalar field rises corresponds to quantum nucleation of a vacuum state represented by $\widetilde{\phi}$. As we will address later, there is a question if there is a generic 'type' of vacuum state as a starting point for the transformation to standard inflation, as given by the 2nd scalar potential system.

The potentials $\widetilde{V}_1$, and $\widetilde{V}_2$ were described in terms of soliton-anti soliton style di quark pairs nucleating in a manner similar in part, for the first potential similar in part to what is observed in instanton physics showing up in density wave current problems, while the second potential is Guth's typical chaotic inflationary cosmology potential dealing with the flatness problem.

Note that this requires that we write $\phi_C$ in Eqn. (43) as an equilibrium value of a true vacuum minimum in Eqn. (44) after quantum tunneling through a barrier. Note that $M_P =$ Planck's mass $\approx 1.2209 \times 10^{19}$ GeV/c$^2$ = $2.176 \times 10^{-8}$ kg

Planck's mass is the mass for which the Schwarzschild radius is equal to the Compton length divided by $\pi$. A Schwarzschild radius is proportional to the mass, with a proportionality constant involving the gravitational constant and the speed of light. The formula for the Schwarzschild radius can be found by setting the escape velocity to the speed of light Furthermore, mass $m << M_P$. Frequently, m is called the mass of the gravitating object. And as a final note, we have that a soliton is a self-reinforcing solitary wave caused by a delicate balance between nonlinear and dispersive effects in the medium. Solitons are found in many physical phenomena, as they arise as the solutions of a widespread class of weakly nonlinear dispersive partial differential equations describing physical systems.
.

## XI. MODELING A FIFTH DIMENSION FOR EMBEDDING FOUR DIMENSIONAL SPACE TIME

We address how to incorporate a more accurate reading of phase evolution and the minimum requirements of phase evolution behavior in a potential system permitting baryogenesis, which imply using a Sundrum fifth-dimension. The fifth-dimension of the Randall-Sundrum brane world is, for $-\pi \leq \theta \leq \pi$, a circle map which is written, with $R$ as the radius of the compact dimension $x_5$ Circle maps were first proposed by Andrey Kolmogorov as a simplified model for driven mechanical rotors (specifically, a free-spinning wheel weakly coupled by a spring to a motor). The circle map equations also describe a simplified model of the phase-locked loop in electronics. We are using a circle map here as a simple way to give a compact geometry to higher dimensional structures which are extremely important in early universe geometries. A closed string can wind around a periodic dimension an integral number of times. Similar to the Kaluza-Klein case they contribute a momentum which goes as p = w R (w=0, 1, 2 ...). The crucial difference here is that this goes the other way with respect to the radius of the compact dimension, R. As now as the compact dimension becomes very small these winding modes are becoming very light! For our purposes, we write our fifth dimension as. [25]

$$x_5 \equiv R \cdot \theta \qquad (58)$$

This fifth dimension $x_5$ also creates an embedding potential structure leading to a complimentary embedded in five dimensions scalar field we model as [25]:

$$\phi(x^\mu, \theta) = \frac{1}{\sqrt{2 \cdot \pi \cdot R}} \cdot \left\{ \phi_0(x) + \sum_{n=1} [\phi_n(x) \cdot \exp(i \cdot n \cdot \theta) + C.C.] \right\} \tag{59}$$

This scaled potential structure will be instrumental in forming a Randall Sundrum effective potential

## XII. RANDALL SUNDRUM EFFECTIVE POTENTIAL

The consequences of the fifth-dimension considered in Eqn. (58) show up in a simple warped compactification involving two branes, i.e., a Planck world brane, and an IR brane. Let's call the brane where gravity is localized the Planck brane The first brane is a four dimensional structure defining the standard model 'universe', whereas the second brane is put in as structure to permit solving the five dimensional Einstein equations. Before proceeding, we need to say what we call the graviton is, in the brane world context. In physics, the graviton is a hypothetical elementary particle that mediates the force of gravity in the framework of quantum field theory. If it exists, the graviton must be massless (because the gravitational force has unlimited range) and must have a spin of 2 (because gravity is a second-rank tensor field). When we refer to string theory, at high energies (processes with energies close or above the Planck scale) because of infinities arising due to quantum effects (in technical terms, gravitation is nonrenormalizable.) gravitons run into serious theoretical difficulties. A localized graviton plus a second brane separated from the brane on which the standard model of particle physics is housed provides a natural solution to the hierarchy problem—the problem of why gravity is so incredibly weak. The strength of gravity depends on location, and away from the Planck brane it is exponentially suppressed. We can think of the brane geometry, in particular the IR brane as equivalent to a needed symmetry to solve a set of equations. This construction permits (assuming K is a constant picked to fit brane world requirements) [25]

$$S_5 = \int d^4 x \cdot \int_{-\pi}^{\pi} d\theta \cdot R \cdot \left\{ \frac{1}{2} \cdot (\partial_M \phi)^2 - \frac{m_5^2}{2} \cdot \phi^2 - K \cdot \phi \cdot [\delta(x_5) + \delta(x_5 - \pi \cdot R)] \right\} \tag{59}$$

Here, what is called $m_5^2$ can be linked to Kaluza Klein "excitations" via (for a number $n > 0$)

$$m_n^2 \equiv \frac{n^2}{R^2} + m_5^2 \tag{60}$$

To build the Kaluza–Klein theory, one picks an invariant metric on the circle $S^1$ that is the fiber of the $U(1)$-bundle of electromagnetism. In this discussion, an *invariant metric* is simply one that is invariant under rotations of the circle. We are using a variant of that construction via Eqn. (8) above. Note that In 1926, Oskar Klein proposed that the fourth spatial dimension is curled up in a circle of very small radius, so that a particle moving a short distance along that axis would return to where it began. The distance a particle can travel before reaching its initial position is said to be the size of the dimension. This extra dimension is a compact set, and the phenomenon of having a space-time with compact dimensions is referred to as compactification in modern geometry.

Now, if we are looking at an addition of a second scalar term of opposite sign, but of equal magnitude, where

$$S_5 = -\int d^4 x \cdot V_{eff}(R_{phys}(x)) \to -\int d^4 x \cdot \widetilde{V}(R_{phys}(x)) \tag{60}$$

We should briefly note what an effective potential is in this situation. [25]

We get

$$\widetilde{V}_{eff}(R_{phys}(x)) = \frac{K^2}{2 \cdot m_5} \cdot \frac{1 + \exp(m_5 \cdot \pi \cdot R_{phys}(x))}{1 - \exp(m_5 \cdot \pi \cdot R_{phys}(x))} + \frac{\widetilde{K}^2}{2 \cdot \widetilde{m}_5} \cdot \frac{1 - \exp(\widetilde{m}_5 \cdot \pi \cdot R_{phys}(x))}{1 + \exp(\widetilde{m}_5 \cdot \pi \cdot R_{phys}(x))} \tag{61}$$

This above system has a metastable vacuum for a given special value of $R_{phys}(x)$. Start with

$$\Psi \propto \exp(-\int d^3 x_{space} d\tau_{Euclidian} L_E) \equiv \exp\left(-\int d^4 x \cdot L_E\right) \tag{62}$$

$$L_E \geq |Q| + \frac{1}{2} \cdot (\tilde{\phi} - \phi_0)^2 \{\ \} \xrightarrow[Q \to 0]{} \frac{1}{2} \cdot (\tilde{\phi} - \phi_0)^2 \cdot \{\ \} \tag{63}$$

Part of the integrand in Eqn. (41) is known as an action integral, $S = \int L \cdot dt$ where L is the Lagrangian of the system. Where as we also are assuming a change to what is known as Euclidean time, via $\tau = i \cdot t$, which has the effect of inverting the potential to emphasize the quantum bounce hypothesis of Sidney Coleman. In that hypothesis, L is the Lagrangian with a vanishing kinetic energy contribution, i.e. $L \to V$, where $V$ is a potential whose graph is 'inverted' by the Euclidian time. Here, the spatial dimension $R_{phys}(x)$ is defined so that

$$\tilde{V}_{eff}(R_{phys}(x)) \approx cont. + (1/2) \cdot (R_{phys}(x) - R_{critical})^2 \propto \tilde{V}_2(\tilde{\phi}) \propto \frac{1}{2} \cdot (\tilde{\phi} - \phi_C)^2 \tag{64}$$

And

$$\{\ \} = 2 \cdot \Delta \cdot E_{gap} \tag{65}$$

We should note that the quantity $\{\ \} = 2 \cdot \Delta \cdot E_{gap}$ referred to above has a shift in minimum energy values between a false vacuum minimum energy value, $E_{\text{false min}}$, and a true vacuum minimum energy $E_{\text{true min}}$, with the difference in energy reflected in Eqn. (65) above.

This requires, if we take this analogy seriously the following identification.

$$\tilde{V}_{eff}(R_{phys}(x)) \approx \textbf{Constant} + \frac{1}{2} \cdot (R_{phys}(x) - R_{critical})^2 \propto V_0 + \frac{m}{2} \cdot [\phi - \varphi_{fluctuations}]^2_{4-dim} \tag{66}$$

So that one can make equivalence between the following statements. These need to be verified via serious analysis.

$$\textbf{Constant} \leftrightarrow V_0 \tag{66a}$$

$$\frac{1}{2} \cdot (R_{phys}(x) - R_{critical})^2 \leftrightarrow \frac{m}{2} \cdot [\phi - \varphi_{fluctuations}]^2_{4-dim} \tag{66b}$$

## XIII. USING OUR BOUND TO THE COSMOLOGICAL CONSTANT

We use our bound to the cosmological constant to obtain a conditional escape of gravitons from an early universe brane. To begin, we present conditions [26] (Leach and Lesame, 2005) for gravitation production. Here $R$ is proportional to the scale factor 'distance'.

$$B^2(R) = \frac{f_k(R)}{R^2} \tag{67}$$

Also there exists an 'impact parameter'

$$b^2 = \frac{E^2}{P^2} \tag{68}$$

This leads to, practically, a condition of 'accessibility' via R so defined with respect to 'bulk dimensions'

$$b \geq B(R) \tag{69}$$

$$f_k(R) = k + \frac{R^2}{l^2} - \frac{\mu}{R^2} \tag{70}$$

Here, k = 0 for flat space, k = -1 for hyperbolic three space, and k = 1 for a three sphere, while an radius of curvature

$$l \equiv \sqrt{\frac{-6}{\Lambda_{5-dim}}} \tag{71}$$

This assumes a negative bulk cosmological constant negative bulk cosmological constant $\Lambda_{5-dim}$ and that $\mu$ is a five dimensional Schwartzshield mass. We also set $R_b(t) = a(t)$. Then we have a maximum effective potential of gravitons defined via

$$B^2(R_t) = \frac{1}{l^2} + \frac{1}{4 \cdot \mu} \tag{72}$$

This leads to a bound with respect to release of a graviton from an anti De Sitter brane [26] (Leach and Lesame, 2005) as

$$b \geq B(R_t) \tag{73}$$

In the language of general relativity, anti de Sitter space is the maximally symmetric, vacuum solution of Einstein's field equation with a negative cosmological constant $\Lambda$ How do we link this to our problem with respect to di quark contributions to a cosmological constant? Here we make several claims.

**Claim 1**: It is possible to redefine $l \equiv \sqrt{\frac{-6}{\Lambda_{5-dim}}}$ as

$$l_{eff} = \sqrt{\left|\frac{6}{\Lambda_{eff}}\right|} \tag{74}$$

**Proof of Claim 1**: There is a way, for finite temperatures for defining a given four-dimensional cosmological constant [27] (Park, Kim,).

We define, via Park's article,

$$k^* = \left(\frac{1}{\text{'AdS curvature}}\right) \tag{75}$$

Park et al write, if we have a 'horizon' temperature term

$$U_T \propto (external \quad temperature) \tag{76}$$

We can define a quantity

$$\varepsilon^* = \frac{U_T^4}{k^*} \tag{77}$$

Then there exists a relationship between a four-dimensional version of the $\Lambda_{eff}$, which may be defined by noting

$$\Lambda_{5-dim} \equiv -3 \cdot \Lambda_{4-dim} \cdot \left(\frac{U_T}{k^{*3}}\right)^{-1} \propto -3 \cdot \Lambda_{4-dim} \cdot \left(\frac{external\ temperature}{k^{*3}}\right)^{-1} \tag{78}$$

So

$$\Lambda_{5-dim} \xrightarrow[external\ temperature \to small]{} \text{Very large value} \tag{79}$$

And set

$$|\Lambda_{5-dim}| = \Lambda_{eff} \tag{80}$$

In working with these values, one should pay attention to how $\cdot \Lambda_{4-dim}$ is defined by Park, et al

$$\cdot \Lambda_{4-dim} = 8 \cdot M_5^3 \cdot k^* \cdot \varepsilon^* \xrightarrow[external\ temperature \to 3\ Kelvin]{} (.0004 eV)^4 \tag{81}$$

Here, I am defining $\Lambda_{eff}$ as being an input from Eqn. (42) to (43) to Eqn (44) above partly due to [27]

$$\Delta \Lambda_{total}|_{effective} = \lambda_{other} + \Delta V$$
$$\xrightarrow[\Delta V \to end\ chaotic\ inflation\ potential]{} \Lambda_{observed} \cong \Lambda_{4-dim}(3\ Kelvin) \tag{82}$$

This is for potential $V_{min}$ being defined via transition between the 1st and the 2nd potentials of Eqn. (43) and Eqn. (44)

$$B_{eff}^2(R_t) = \frac{1}{l_{eff}^2} + \frac{1}{4 \cdot \mu} \tag{83}$$

**Claim 2**

$R_b(t) = a(t)$ Ceases to be definable for times $t \leq t_P$ where the upper bound to the time limit is in terms of Planck time and in fact the entire idea of a de Sitter metric is not definable in such a physical regime.

**Claim 3**

Eqn. (43) has a 1st potential which tends to be for a di quark nucleation procedure which just before a defined Planck's time $t_P$. But that the cosmological constant was prior to time $t_P$ likely far higher, perhaps in between the values of the observed cosmological constant of today, and the QCD tabulated cosmological constant which was / is $10^{120}$ time greater. I.e.

$$b^2 \geq B_{eff}^2(R_t) = \frac{1}{l_{eff}^2} + \frac{1}{4 \cdot \mu} \tag{84}$$

With furthermore

$$\left.\frac{1}{l_{eff}^2}\right|_{t \leq t_P} >> \left.\frac{1}{l_{eff}^2}\right|_{t \equiv t_P + \Delta(time)} \tag{85}$$

So then that there would be a great release of gravitons at or about time $t_P$.

**Claim 4**.

Few gravitons would be produced significantly after time time $t_P$.

Proof of claim 4

This comes as a result of temperature changes after the initiation of inflation and changes in value of

$$(\Delta l_{eff})^{-1} = \left(\sqrt{\left|\frac{6}{\Lambda_{eff}}\right|}\right)^{-1} \propto \Delta\left(external \quad temperature\right) \tag{86}$$

The existence and evolution of a scale factor ceasing to be definable as presented in **claim 2** is due to the construction of typical GR metrics breaking down completely when one has a strongly curved space, which is what we would expect in the first instant of less than Planck time evolution of the nucleation of a new universe.

So then that there would be a great release of gravitons at or about time $t_P$.

## XIV. BRANE WORLD AND DI QUARK LEAST ACTION INTEGRALS

Now for the question we are raising: Can we state the following for initial conditions of a nucleating universe?

$$S_5 = -\int d^4x \cdot \tilde{V}_{eff}\left(R_{phys}(x)\right) \propto \left(-\int d^3x_{space} d\tau_{Euclidian} L_E\right) \equiv \left(-\int d^4x \cdot L_E\right) \tag{87}$$

This leads to ask whether we should instead look at what can be done with S-S' instanton physics and the Bogolmyi inequality, in order to take into account baryogenesis. In physical cosmology, baryogenesis is the generic term for hypothetical physical processes that produced an asymmetry between baryons and anti-baryons in the very early universe, resulting in the substantial amounts of residual matter that comprises the universe today. $L_E$ Is almost the same as Eqn. (41) above and requires elaboration of Eqn. (87) above. We should think of Eqn. (87) happening in the Planck brane mentioned above. Keep in mind that there are many baryogenesis theories in existence, The fundamental difference between baryogenesis theories is the description of the interactions between fundamental particles, and what we are doing with di quarks is actually one of the simpler ones.

## XV. DI QUARK POTENTIAL SYSTEMS AND THE WHEELER DE-WITT EQUATION

Abbay Ashtekar's quantum bounce [6, 7] gives a discrete version of the Wheeler De Witt equation; we begin with [28]

$$\psi_\mu(\phi) \equiv \psi_\mu \cdot \exp(\alpha_\mu \cdot \phi^2) \tag{88}$$

As well as energy term As well as an energy term

$$E_\mu = \sqrt{A_\mu \cdot B_\mu} \cdot m \cdot \hbar \tag{89}$$

$$\alpha_\mu = \sqrt{B_\mu / A_\mu} \cdot m \cdot \hbar \tag{90}$$

This is for a 'cosmic' Schrodinger equation as given by

$$\tilde{\tilde{H}} \cdot \psi_\mu(\phi) = E_\mu(\phi) \tag{91}$$

This has $V_\mu$ is the eignvalue of a so called volume operator. So:

$$A_\mu = \frac{4 \cdot m_{pl}}{9 \cdot l_{pl}^9} \cdot \left(V_{\mu+\mu_0}^{1/2} - V_{\mu-\mu_0}^{1/2}\right)^6 \tag{92}$$

And

$$B_\mu = \frac{m_{pl}}{l_{pl}^3} \cdot (V_\mu) \tag{93}$$

Key to doing this though is to work with a momentum basis for which we have

$$\hat{p}_\iota |\mu\rangle = \frac{8 \cdot \pi \cdot \gamma \cdot l_{PL}^2}{6} \cdot \mu |\mu\rangle \tag{94}$$

With the advent of this re definition of momentum we are seeing what Ashtekar works with as a simplistic structure with a revision of the differential equation assumed in Wheeler – De Witt theory to a form characterized by [6,7] Eqn (102) above. This is akin to putting in the sympletic structure alluded to by Ashkekar [6, 7]. This is a generalization of what, Alfredo B. Henrique's wrote as a way in which one can obtain a Wheeler De Witt equation based upon

$$\tilde{\tilde{H}} \cdot \Psi(\phi) = \left[\frac{1}{2} \cdot \left(A_\mu \cdot p_\phi^2 + B_\mu \cdot m^2 \cdot \phi^2\right) \cdot \Psi(\phi)\right] \tag{95}$$

Using a momentum operator as give by

$$\hat{p}_\iota = -i \cdot \hbar \cdot \frac{\partial}{\partial \cdot \phi} \tag{96}$$

Ashtekar [6, 7] works with as a simplistic structure with a revision of the differential equation assumed in Wheeler-De Witt theory to a form characterized by

$$\frac{\partial^2}{\partial \phi^2} \cdot \Psi \equiv - \Theta \cdot \Psi \tag{97}$$

$\Theta$ In this situation is such that

$$\Theta \neq \Theta(\phi) \tag{98}$$

This will lead to $\Psi$ having roughly the form alluded to in Eqn. (87), which in early universe geometry will eventually no longer be $L^P$, but will have a discrete geometry. This may permit an early universe 'quantum bounce' and an outline of an earlier universe collapsing , and then being recycled to match present day inflationary expansion parameters. The main idea behind the quantum theory of a (big) quantum bounce is that, as density approaches infinity, so the behavior of the quantum foam changes. The foam is a qualitative description of the turbulence that the phenomenon creates at extremely small distances of the order of the Planck length. Here $V_\mu$ is the Eigen value of a so called volume operator and we need to keep in mid that the main point made above, is that a potential operator based upon a quadratic term leads to a Gaussian wave function with an exponential similarly dependent upon a quadratic $\phi^2$ exponent., and more importantly this $\Theta$ is a difference operator, allowing for a treatment of the scalar field as an 'emergent time', or 'internal time' so that one can set up a wave functional built about a Gaussian wave functional defined via

$$\max \widetilde{\Psi}(k) = \widetilde{\Psi}(k)\big|_{k \equiv k^*} \tag{99}$$

This is for a crucial 'momentum' value

$$p_\phi^* = -\left(\sqrt{16 \cdot \pi \cdot G \cdot \hbar^2 / 3}\right) \cdot k^* \tag{100}$$

And

$$\phi^* = -\sqrt{3/16 \cdot \pi G} \cdot \ln|\mu^*| + \phi_0 \tag{101}$$

Which leads to, for an initial point in 'trajectory space' given by the following relation $(\mu^*, \phi_0) =$ (initial degrees of freedom [dimensionless number] ~'eignvalue of 'momentum', initial 'emergent time ' ) So that if we consider eignfunctions of the De Witt (difference) operator, as contributing toward

$$e_k^s(\mu) = (1/\sqrt{2}) \cdot [e_k(\mu) + e_k(-\mu)] \tag{102}$$

With each $e_k(\mu)$ an eignfunction of Eqn. (94) above, we have a potentially numerically treatable early universe wave functional data set which can be written as

$$\Psi(\mu, \phi) = \int_{-\infty}^{\infty} dk \cdot \widetilde{\Psi}(k) \cdot e_k^s(\mu) \cdot \exp[i\omega(k) \cdot \phi] \tag{103}$$

## XVI. DETECTING GRAVITONS AS SPIN 2 OBJECTS WITH AVAILABLE TECHNOLOGY

Let us now briefly review what we can say now about standard graviton detection schemes. As mentioned earlier, T. Rothman [29] states that the Dyson seriously doubts we will be able to detect gravitons via present detector technology. The conundrum is as follows, namely if one defines the criterion for observing a graviton as

$$\frac{f_\gamma \cdot \sigma}{4 \cdot \pi} \cdot \left(\frac{\alpha}{\alpha_g}\right)^{3/2} \cdot \frac{M_s}{R^2} \cdot \frac{1}{\varepsilon_\gamma} \geq 1 \tag{104}$$

Here,

$$f_\gamma = \frac{\breve{L}_\gamma}{\breve{L}} \tag{105}$$

This has $\frac{\breve{L}_\gamma}{\breve{L}}$ a graviton sources luminosity divided by total luminosity and $R$ as the distance from the graviton source, to a detector. Furthermore, $\alpha = e^2/\hbar$ and $\alpha_g = Gm_p^2/\hbar$ are constants, while $\varepsilon_\gamma$ is the graviton potential energy. As stated in the manuscript, the problem then becomes determining a cross section $\sigma$ for a graviton production process and $f_\gamma = \frac{\breve{L}_\gamma}{\breve{L}}$. The existence of branes is relevant to graviton production. Here, $\breve{L}_\gamma$ is luminosity of graviton producing process $\geq 7.9 \times 10$ to the 14th ergs/s, while $\breve{L}$ is the general background luminosity which is usually much less than $\breve{L}_\gamma$. At best, we usually can set $f_\gamma = .2$, which does not help us very much. That means we need to look else where than the usual processes to get satisfaction for Graviton detection. This in part is why we are looking at relic graviton production for early universe models, usually detectable via the criteria developed for white dwarf stars of one graviton for $10^{13-14}$ neutrinos [30]

We should state that we will generally be referring to a cross section which is frequently the size of the square of Planck's length $l_P$ which means we really have problems in detection, if the luminosity is so low. An upper bound to the cross section $\sigma$ for a graviton production process $\approx 1/M$ with $M$ being written with respect to the Planck's scale in 4+n dimensions which we can set as equal to $\left(M_P^2/\hat{V}_n\right)^{1/2\_n}$, and this is using a very small value of $\hat{V}_n$ as a compactified early universe extra dimension 'square' volume $\approx$ 10-15 mm per side. All this geometry is congruent with respect to the sympletic geometry structure alluded to for the wave functional as typically given by Eqn. (102) above.

This in itself would permit confirmation of if or not a quantum bounce condition existed in early universe geometry, according to what Ashtekar's two articles predict. In addition it also corrects for another problem. Prior to brane theory we had a too crude model. Why? When we assume that a radius of an early universe, assuming setting the speed of light to one, is of the order of magnitude 3 $\Delta t$ with the change in time alluded to of the order of magnitude of Planck's time $t_P$. So we face a rapidly changing volume that is heavily dependent upon a first order phase transition, as affected by a change in the degrees of freedom given by $(\Delta N(T))_P$. Without gravitons and brane world structure, such a model is insufficient to account for dark matter production and fails to even account for Baryogenesis. It also will lead to new graviton detection equipment re configuration well beyond the scope of falsifiable models configured along the lines of simple phase transitions given for spatial volumes (assuming c = 1) of the form [31]

$$\Delta t \cong t_P \propto \frac{1}{4\pi} \cdot \sqrt{\frac{45}{\pi \cdot (\Delta N(T))_P}} \cdot \left(\frac{M_p}{T^2}\right) \tag{106}$$

This creates problems, so we look for other ways to get what we want. Grushchuk writes that the energy density of relic gravitons is expressible as [32]

$$\varepsilon(v) = \frac{\pi}{(2\cdot\pi)^4 a(t)^4} \cdot H_i^2 \cdot H_f^2 \cdot a(t)_f^4 \tag{107}$$

Where the subscripts $i$ and $f$ refer to initial and final states of the scale factor, and Hubble parameter. This expression though is meaningless in situations when we do not have enough data to define either the scale factor, or Hubble parameter at the onset of inflation. How can we tie in with the Gaussian wave functional given in Eqn (99) and defined as an input into the data used to specify Ashtekar's quantum bounce? Here, we look at appropriate choices for an optimum momentum value for specifying a high level of graviton production. If gravitons are, indeed, for dark energy, as opposed to dark matter, without mass, we can use, to first approximation something similar to using the zeroth component of momentum $p^0 = E(energy)/c$, calling $E(energy) \equiv \varepsilon(v) \cdot$ (initial nucleation volume), we can read off from Eqn. (98), 'pre inflationary' universe values for the k values of Eqn. (99)can be obtained, with an optimal value selected. This is equivalent to using to first approximation the following. The absolute value of $k^*$, which we call $|k^*|$ is

$$|k^*| = \sqrt{3/16 \cdot \pi \cdot G \cdot \hbar^2} \cdot \left(\varepsilon(v) \cdot \left(\text{initial nucleation volume}\right)/c\right) \tag{108}$$

An appropriate value for a Gaussian representation of an instanton awaits more detailed study. But for whatever it is worth we can refer to the known spaleraton value for a multi dimensional instanton via the following procedure. We wish to have a finite time for the emergence of this instanton from a pre inflation state.

If we have this, we are well on our way toward fixing a range of values for $\omega 2 < \omega(net) < \omega 1$, which in turn will help us define

$$\varepsilon(v) \cdot \left(\text{initial volume}\right) \approx \hbar \cdot \omega(net) \equiv p^* \cdot c \tag{109}$$

In order to use Eqn. (108) to get a value for $k^*$. This value for $k^*$ can then is used to construct a Gaussian wave functional about $k^*$ of the form, as an anzatz. To put into Eqn. (102) above.

$$\Psi(k) \approx \frac{1}{Value} \cdot \exp\left(-c_2 \cdot (k - k^*)^2\right) \tag{110}$$

If so, then, most likely, the question we need to ask though is the temperature of the 'pre inflationary' universe and its link to graviton production. This will be because the relic graviton production would be occurring before the nucleation of a scalar field. We claim, as beforehand that this temperature would be initially quite low, as given by the two University of Chicago articles, but then rising to a value at or near $10^{12}$ degrees Kelvin after the dissolving of the axion wall contribution given in the dominant value of Eqn. (43) leading to Eqn (44) for a chaotic inflationary potential.

# XVII. TIE IN WITH ANSWERING GUTHS QUESTION ABOUT THE EXISTENCE OF A PREFERRED VACUUM NUCLEATION STATE?

First of all, this is separate from the question of the existence of a scale factor. We are assuming that the scale factor would exist for cosmological times of the order of magnitude of Planck's time interval. This is also assuming that the nucleation of the favored vacuum state would commence for values of a scalar potential in line with conditions leading to the formation of Eqn. (43) above. Having said that, let us commence looking at what suffices to initiate chaotic inflation? The change in the cosmological expansion scale factors so alluded to due to changes from Eqn. 42 to Eqn. 44 as a factor, but we also need to look to if or not we need the slow roll condition. We argue that we do, and that models, like Ghost inflation, which purport to explain cosmic evolution without inflation run up against serious problems. And this leads to considering if or not we have a preferred initial nucleation state.

This is akin to making the following assertion. Namely that at the start of a new universe that the relationship given below ceases to hold, namely [27]:

$$\left. \frac{H^2}{|\dot{\phi}_{cl}|} \right|_{t \equiv t_P} \neq \frac{\delta\rho}{\rho} \Rightarrow \text{(Scalar) density perturbations are NOT of order } O(1) \text{ at time } t_P \quad (111)$$

And that one needs a new starting point for pre inflationary cosmological models. To get this start, we shall try to ascertain if or not a favored vacuum state actually exists. We claim it does. The action given by the following structure $S_5 = -\int d^4x \cdot \tilde{V}_{eff}(R_{phys}(x)) \propto (-\int d^3 x_{space} d\tau_{Euclidian} L_E) \equiv (-\int d^4 x \cdot L_E)$ is in tandem with writing along the lines of the quantum bounce as given by Sidney Coleman [33], and treating $S_5|_{non-Euclidian\ time}$ as a phase factor

$$\psi_{wave\ function\ of\ the\ universe} \sim \exp(-S_5) \equiv \exp(i \cdot S_5|_{non-Euclidian\ time}) \quad (112)$$

via the following quantum mechanical theorem: We have a path way to a generic behavior of an ensemble of wave functionals with equivalent phase evolution behavior. Taking into account the quantum mechanical theorem stated as mentioned below:

In quantum theory for example it is well known that you may change the phase of the wave functions by an arbitrary amount without altering any the physical content or structure of the theory, provided that you change the all wave functions in the same way, everywhere in space. We are doing the same thing here with respect to:

$$\text{Wave function} \sim \exp(-i \cdot \text{integral (effective potential)}) \quad (113)$$

That condition of changing the ensemble of wave functions in the same way will force the simplest form of construction, of a vacuum state consistent with respect to Eqn. (113) above as a favored initial starting point for forming a phase evolution consistent with one favored vacuum state.

The upshot of having a favored vacuum state is that we can then ascertain if or not we have a formation of a quantum bounce in line with Ashtekar, A., Pawlowski, T. and Singh, P [6,7] (2005)suppositions given in the modification of the Wheeler – De Witt equation given in their Phys Rev. D article. This in its own way would lead to investigating the feasibility of a prior universe providing us with initial input of energy needed to stimulate relic graviton production on the scale so visualized, as well as reconciling the initial low temperature state visualized by a gravity dominated early universe with low entropy, with the peaks of temperature given at the onset of Guth inflationary potential cosmology. [27]

# XVIII. SIMILARITIES/DIFFERENCES WITH GHOST INFLATION, AND INQUIRES AS TO THE ROLE OF CONDENSATES FOR INITIAL VACUUM STATES

Arkani-Hamed [34] recently has used the Ghost inflation paradigm to eliminate using slow roll as a way to initiate inflation with far lower energy scales than is usually associated with standard inflation models. His prediction does, which we disagree with, postulate far fewer gravity waves / relic gravitons than is associated with standard models of inflation. We postulate MORE, rather than less assumed relic graviton production. However, we also think that his analysis makes many cogent points which we will enumerate here, which are pertinent to the initial condensate nucleation, which we will put forward here.

Standard slow roll is premised upon the following quantum fluctuation assumptions [27]:

1) Quantum fluctuations are important on small scales, if and only if one is working with a static space time (i.e. no expanding universe)
2) For inflating space times, quantum fluctuations are 'expanded' to be congruent in magnitude with classical sizes ( classical fluctuations)
3) Simple random walk picture: In each time interval of $\Delta t \equiv H^{-1}$, the average field $\phi$ receives an increment with root means squared, of $\Delta \phi_{qu} = \frac{H}{2 \cdot \pi}$. This increment is super imposed upon the classical motion, which is downward.
4) Quantum fluctuations are equally likely to move field $\phi$ 'up or down' the well of a 'harmonic' style potential.

Those who read the presentation should note the conclusion which is something which raises serious questions: i.e.

5) In early universe geometry, the probability of an upward (quantum)fluctuation exceeds $1/e^3 \approx 1/20$ if

$$\Delta \phi_{quantum} \approx \frac{H}{2 \cdot \pi} > 0.61 \times \left| \phi_{classical\ value} \right| \cdot H^{-1} \Leftrightarrow \frac{H^2}{\left| \phi_{classical\ value} \right|} > 3.8 \tag{114}$$

But

$$\frac{H^2}{\left| \phi_{classical} \right|} \sim \frac{\delta \rho}{\rho} \quad \text{(Scalar) density perturbations are of order } O(1) \tag{115}$$

In addition, we have that if we look at

$$\left( \frac{V_\phi}{V} \right)^2 << 1 \tag{116}$$

We find that $\widetilde{V}_1$ of Eqn. (43) fits this requirement for small $\phi$ values, but is inconsistent with respect to

$$\left| \left( \frac{V_{\phi\phi}}{V} \right) \right| << 1 \tag{117}$$

However, if we work with $\widetilde{V}_2$ of Eqn. (44) that both of these conditions would be amply satisfied. We can either do two things. First of all state that $\widetilde{V}_1$ of Eqn. (42) and Eqn. (43) is such a fleeting instant of nucleated time, that the slow roll condition does not hold, and consider that the de facto history as we can manage it of Cosmological evolution is after a time $\Delta t \approx t_P$ in which we consider only $\widetilde{V}_2$ of Eqn. (44). If we consider though that the initial phases of nucleation as we postulate are our candidate for relic graviton production we need to question how feasible making the assumption so out lined as to ignoring Eqn. (117) are for such a short instant of time.

Enter in Arkani-Hamed's ghost inflation paper. He configures the evolution of de Sitter phases via a ghost scalar field $\hat{\phi}$ condenses in a background with a non zero velocity along the lines of (assuming M is a generic mass term)

$$\langle \dot{\hat{\phi}} \rangle = M^2 \Rightarrow \langle \hat{\phi} \rangle = M^2 t \qquad (118)$$

This leads to a density fluctuation along the lines of, assuming an upper bound of $M \leq 10 MeV$

$$\left. \frac{\delta \rho}{\rho} \right|_{Ghost} \approx \left( \frac{H}{M} \right)^{5/4} \qquad (119)$$

As opposed to, when $\in$ is a slow roll parameter proportional to $(V'/V)^2$

$$\left. \frac{\delta \rho}{\rho} \right|_{Ordinary\ inf} \approx \left( \frac{H}{m_p \cdot \sqrt{\in}} \right) \qquad (120)$$

The ghost inflation paradigm so outlined postulates that there exists a maximum energy scale of the order of $V_0 \sim (1000 TeV)^4$ which allegedly rules out relic gravitons. The model so outlined here, which we are working with assumes a massive relic graviton production surge. So it appears that we cannot ignore some variant of inflation. We need to do additional investigations as to if or not it is realistic to suppose that time restrictions below Planck time are enough to lead to a 'temporary' violation of Eqn. (117)

We thereby will from now on stick to our model which appears to give criteria for graviton production. But Arkani-Hamed's ghost inflation [34] if true would probably eliminate the feasibility of graviton space travel systems. We have outlined initial universe conditions which if replicated would allow them to exist, and which could be used for space travel. How now do we stick this to gravity waves? Note that for gravity fields, we have an analogous procedure as to how magnetic fields form. I.e.

There are three main contributions to the generation of magnetic fields (i) the baryon-photon slip term, (ii) the vorticity difference term, and (iii) the anisotropic pressure term. These terms are derived from the fact that electrons are pushed by photons through Compton scattering when velocity differences exist between them or when there is anisotropic pressure from photons. This in a word is due to early universe turbulence and the growth of structure. But it is still not going to get us past the red shift barrier as of Z = 1000 or so already mentioned. Note that as stated by Alexander D. Dolgov et al [35]

"Periods of cosmic turbulence may have left a detectable relic in the form of stochastic backgrounds of gravitational waves" and that we are looking at traces of neutrino inhomogeneous diffusion and a first order phase transition to model the spectrum of gravity waves one may observe. Then we have to recall that we have one graviton for $10^{13-14}$ neutrinos. This means we need a huge first order phase transition, which Arkani-Hamed's ghost inflation does not provide.

# XIX. GRAVITON SPACE PROPULSION SYSTEMS

We need to understand what is required for realistic space propulsion. To do this, we need to refer to a power spectrum value which can be associated with the emission of a graviton. Fortunately, the literature contains a working expression as to power generation for a graviton being produced for a rod spinning at a frequency per second $\omega$, which is by Fontana (2005) [36] at a STAIF new frontiers meeting, which allegedly gives for a rod of length $\widehat{L}$ and of mass m a formula for graviton production power,

$$P(power) = 2 \cdot \frac{m_{graviton}^2 \cdot \widehat{L}^4 \cdot \omega_{net}^6}{45 \cdot (c^5 \cdot G)} \tag{121}$$

The point is though that we need to say something about the contribution of frequency needs to be understood as a mechanical analogue to the brute mechanics of graviton production. For the sake of understanding this, we can view the frequency $\omega_{net}$ as an input from an energy value, with graviton production number (in terms of energy) as given approximately via an integration of Eqn. (51) above, $\widehat{L} \propto l_P$ mass $m_{graviton} \propto 10^{-60} kg$. This crude estimate of graviton power production will be considerably refined via numerical techniques in the coming months. It also depends upon a HUGE **number** of relic gravitons being produced, due to the temperature variation so proposed.

One can see the results of integrating

$$\langle n(\omega) \rangle = \frac{1}{\omega(net\ value)} \int_{\omega 1}^{\omega 2} \frac{\omega^2 d\omega}{\pi^2} \cdot \left[ \exp\left( \frac{2 \cdot \pi \cdot \hbar \cdot \omega}{\bar{k}T} \right) - 1 \right]^{-1} \tag{122}$$

And then one can set, if $\hbar \equiv 1$, and a normalized 'energy input' as $E_{eff} \equiv \langle n(\omega) \rangle \cdot \omega \equiv \omega_{eff}$

This will lead to the following table of results, with $T^*$ being an initial thermal background temperature of the pre inflationary universe condition

| N1=1.794 E-6 for $Temp = T^*$ | Power = 0 |
| --- | --- |
| N2=1.133 E-4 for $Temp = 2T^*$ | Power = 0 |
| N3= 7.872 E+21 for $Temp = 3T^*$ | Power = 1.058 E+16 |
| N4= 3.612E+16 for $Temp = 4T^*$ | Power $\cong$ very small value |
| N5= 4.205E-3 for $Temp = 5T^*$ | Power= 0 |

The outcome is that there is a distinct power spike associated with Eqn. 121 and Eqn. 122, which is congruent with a Relic graviton burst, assuming when one does this that the back ground in the initial inflation state causes a thermal heat up of the axion wall 'material' due to a thermal input from a prior universe quantum bounce

# XX. BO FENG'S ANALYSIS OF CPT VIOLATIONS DUE TO THE PHYSICS OF QUINTESSENCE FIELDS

Bo Feng et al. [10] as of 2006 presented a phenomenological effective Lagrangian to provide an argument as to possible CPT violations in the early universe. Their model stated, specifying a non zero value to $(\partial_0 \phi)$ where $(\partial_0 \phi) \neq 0$ is implicitly assumed in Eqns. (42) to (44) above.

$$L_{eff} \sim (\partial_u \phi) \cdot A_v \cdot \begin{pmatrix} 0 & -B_x & -B_y & B_z \\ B_x & 0 & E_z & -E_y \\ B_y & -E_z & 0 & E_X \\ B_z & E_y & -E_x & 0 \end{pmatrix} \quad \ldots\ldots\ldots\ldots (123)$$

We are providing a necessary set of conditions for $(\partial_0 \phi) \neq 0$ via a first order early universe phase transition. Bo Feng and his research colleagues gave plenty of evidence via CMB style arguments as to the existence of non zero 'electric' and 'magnetic' field contributions to the left hand side of Eqn. (123) above. In addition they state that TC and GC correlation power spectra require the violation of parity, a supposition which seems to be supported via a change of a polarization plane $\Delta\alpha$ via having incoming photons having their polarization vector of each incoming photon rotated by an angle $\Delta\alpha$ for indicating what they call 'cosmological birefringence' [10]. The difference between their results and ours lies in the CMB limiting value of z as no larger than about 1000, i.e. as of when the universe was 100,000 times younger than today. This indicates a non zero $(\partial_0 \phi) \neq 0$, but in itself does not provide dynamics as to the early evolution of a quintessence scalar field, which our manuscript out lines above. The work done by Bo Feng is still limited by the zero probability of observing relic photon production at the onset of inflation, at or about a Planck time instance due to the reasons Weinberg outlined in his 1977 tome [8] on cosmology, whereas what we did was to give predictions as to how quintessence in scalar fields evolves before the z = 1000 red shift barrier.

# XXI. DYNAMICS OF AXION INTERACTION WITH BARYONIC MATTER, VIA QUINTESSENCE SCALAR FIELD

This discussion is modeled on an earlier paper on Quintessence and spontaneous Leptogenesis (baryogenesis) by M. Li, X. Wang, B.Feng, and Z. Zhang [13] which gave an effective Lagrangian, and an equation of 'motion' for quintessence which yielded four significant cases for our perusal. The last case, giving a way to reconcile the influx of thermal energy of a quantum bounce into an axion dominated initial cosmology, which lead to dissolution of the excess axion 'mass'. This final reduction of axion 'mass' via temperature variation leads to the Guth style chaotic inflationary regime.

Let us now look at a different effective Lagrangian which has some similarities to B. Feng's effective Lagrangian, equation 123, but which leads to our equations of motion for Quintessence scalar fields, assuming as was in Eqn. 123 that specifying a non zero value to $(\partial_0 \phi)$ where $(\partial_0 \phi) \neq 0$ is implicitly assumed in Eqn. 42 to Eqn. 44 i.e.

$$L_{eff} \propto \frac{\tilde{c}}{M} \cdot (\partial_\mu \phi) \cdot J^\mu \quad \ldots\ldots\ldots\ldots\ldots\ldots\ldots\ldots\ldots (124)$$

What will be significant will be the constant, $\tilde{c}$ which is the strength of interaction between a quintessence scalar field and baryonic matter. M in the denominator is a mass scale which can be either $M \equiv M_{planck}$, or $M \equiv M_{GUT}$ is not so important to our discussion, and $J^\mu$ is in reference to a baryonic 'current'. The main contribution to our analysis this paper gives us is in their quintessence 'equation of motion' which we will present, next. Note, that what we are calling $g_b$ is the degrees of freedom of baryonic states of matter, and $T$ is a back ground temperature w.r.t. early universe conditions. $H \cong 1/t(time)$ is the Hubble parameter, with time $t \propto O(t_P)$, i.e. time on the order of Planck's time, or in some cases much smaller than that.

$$\ddot{\phi} \cdot \left[1 + \frac{\tilde{c}}{M^2} \cdot \frac{T^2}{6} \cdot g_b \right] + 3 \cdot H \cdot \dot{\phi} \cdot \left[1 + \frac{\tilde{c}}{M^2} \cdot \frac{T^2}{6} \cdot g_b \right] + \left(\frac{\partial V_{axion-contri}}{\partial \phi}\right) \cong 0 \quad (125)$$

Here I am making the following assumption about the axion contribution scalar potential system

$$V_{axion-contri} \equiv f[m_{axion}(T)] \cdot (1 - \cos(\phi)) + \frac{m^2}{2} \cdot (\phi - \phi_C)^2 \quad (126)$$

For low temperatures, we can assume that prior to inflation, as given by Carroll and Chen [2] we have for $t << t_P$

$$f[m_{axion}(T)]_{T \approx 2^0 \, Kelvin} \propto O((50-100) \cdot m^2) \quad (127)$$

And that right at the point where we have a thermal input with back ground temperatures at or greater than $10^{12} \, Kelvin$ we are observing for $0 < \varepsilon^+ << 1$ and times $t << t_P$

$$f[m_{axion}(T)]_{T \approx 10^{12} \, Kelvin} \propto O((\varepsilon^+) \cdot m^2) \quad (128)$$

This entails having at high enough temperatures

$$V_{axion-contri}|_{T > 10^{12} \, Kelvin} \cong \frac{m^2}{2} \cdot (\phi - \phi_C)^2 \quad \ldots\ldots\ldots\ldots\ldots\ldots\ldots\ldots (129)$$

Let us now review the four cases so mentioned and to use them to analyze new physics

**CASE I:**

Temperature T very small, a.k.a. Carroll and Chen's suppositions (also see Penrose's version of the Jeans inequality) and time less than $t_P$. This is the slow roll case, which is also true when we get to time $>> t_P$

$$\ddot{\phi} \cdot \left[1 + \frac{\tilde{c}}{M^2} \cdot \frac{T^2}{6} \cdot g_b \right] + 3 \cdot H \cdot \dot{\phi} \cdot \left[1 + \frac{\tilde{c}}{M^2} \cdot \frac{T^2}{6} \cdot g_b \right] + \left(\frac{\partial V_{axion-contri}}{\partial \phi}\right)$$

$$\xrightarrow[T \to 0^+]{} 3 \cdot H \cdot \dot{\phi} \cdot + \left(\frac{\partial V_{axion-contri}}{\partial \phi}\right) \cong 0 \quad (130)$$

**CASE II:**

Temperature T very large and time in the neighborhood of $t_P$. This is NOT the slow roll case, and has $H \propto 1/t_P$. Note, which is important that the constant $\tilde{c}$ is not specified to be a small quantity

$$\ddot{\phi} \cdot \left[1 + \frac{\tilde{c}}{M^2} \cdot \frac{T^2}{6} \cdot g_b\right] + 3 \cdot H \cdot \dot{\phi} \cdot \left[1 + \frac{\tilde{c}}{M^2} \cdot \frac{T^2}{6} \cdot g_b\right] + \left(\frac{\partial V_{axion-contri}}{\partial \phi}\right)$$

$$\xrightarrow[T \to 10^{12} \, Kelvin]{} \ddot{\phi} + 3 \cdot H \cdot \dot{\phi} + \left(\frac{\tilde{c}}{M^2} \cdot \frac{T^2}{6} \cdot g_b\right)^{-1} \cdot \left(\frac{\partial V_{axion-contri}}{\partial \phi} = m^2 \cdot (\phi - \phi_C)\right) \cong 0 \tag{131}$$

We then get a general, and a particular solution with $\phi_{general} \propto \exp(p \cdot t)$, $\phi_{particular} \equiv \phi_C$, $\phi_{Total} = \phi_{general} + \phi_{particular}$,

$$p^2 + 3 \cdot H \cdot p + \left(\frac{\tilde{c}}{M^2} \cdot \frac{T^2}{6} \cdot g_b\right)^{-1} \cdot (m^2) \cong 0$$

$$\to p \cong \left[-\frac{3H}{2} \cdot \left[2 - 4 \cdot \frac{m^2 \cdot M^2}{T^2 \cdot c \cdot g_b H}\right], -\left(6 \cdot \frac{m^2 \cdot M^2}{T^2 \cdot c \cdot g_b}\right) \approx -\varepsilon^+\right] \equiv [p_1, p_2] \tag{132}$$

$$\Rightarrow \phi_{general} \cong c_1 \cdot \exp(-|p_1| \cdot t) + c_2 \cdot \exp(-(|p_2| \approx \varepsilon^+) \cdot t)$$

$$\phi_{Total} = \phi_{general} + \phi_{particular} \cong \phi_C + \varepsilon_1 \cdot \phi_{initial \ value} + H.O.T. \text{ , where } \varepsilon_1 < 1 \tag{133}$$

**CASE III:**

Temperature T very large and time in the neighborhood of $t_P$. This is NOT the slow roll case, and has $H \propto 1/t_P$. Note, which is important that the constant $c$ IS specified to be a small quantity. We get much the same analysis as before except the higher order terms (H.O.T.) do not factor in

$$\phi_{Total} = \phi_{general} + \phi_{particular} \cong \phi_C + \varepsilon_1 \cdot \phi_{initial \ value} \text{ , where } \varepsilon_1 < 1 \tag{134}$$

**Case IV:**

Temperature T not necessarily large but on the way of becoming large valued, so the axion mass is not negligible, YET, and time in the neighborhood of $t_P$. This is NOT the slow roll case, and has $H > H_{t=t_P} \propto 1/t_P$. Begin with making the following approximation to the Axion dominated effective potential

$$V_{axion-contri} \equiv f[m_{axion}(T)] \cdot (1 - \cos(\phi)) + \frac{m^2}{2} \cdot (\phi - \phi_C)^2 \Rightarrow$$

$$\left(\frac{\partial V_{axion-contri}}{\partial \phi}\right) \xrightarrow[Temperature \ getting \ larger]{} \tag{135}$$

$$f[m_{axion}(T)] \cdot \frac{\phi^5}{125} - f[m_{axion}(T)] \cdot \frac{\phi^3}{6} + \left[(m^2 + f[m_{axion}(T)]) \cdot \phi - m^2 \phi\right]$$

Then we obtain

$$\ddot{\phi} + 3 \cdot H \cdot \dot{\phi} + \left( \frac{\tilde{c}}{M^2} \cdot \frac{T^2}{6} \cdot g_b \right)^{-1} \cdot \left( \frac{\partial V_{axion-contri}}{\partial \phi} \right) \cong 0 \tag{136}$$

This will lead to as the temperature rises we get that the general solution has definite character as follows

$$p^2 + 3 \cdot H \cdot p + \left( \frac{\tilde{c}}{M^2} \cdot \frac{T^2}{6} \cdot g_b \right)^{-1} \cdot (m^2) \cong 0$$

$$\rightarrow p \cong \left[ -\frac{3H}{2} \cdot \left[ 1 \pm \sqrt{1 - \frac{6 \cdot M^2}{3 \cdot T^2 \cdot c \cdot g_b H} \cdot (m^2 + f[m_{axion}(T)])} \right] \right] \equiv [p_1, p_2]$$

$$\Rightarrow \phi_{general} \cong c_1 \cdot \exp(p_1 \cdot t) + c_2 \cdot \exp(p_2 \cdot t) \tag{137}$$

$$\propto [\phi(real) + i \cdot \phi(imaginary)] \; iff \; [m_{axion}(T)] \; large$$

$$\propto [\phi(real)] \; iff \; [m_{axion}(T)] \; small$$

The upshot is, that for large, but shrinking axion mass contributions we have a cyclical oscillatory system, which breaks down and becomes a real field if the axion mass disappears.

## XXII. CONCLUSION

So far, we have tried to reconcile the following. First, the Chaplygin Gas model is congruent with string theory only if we have the power coefficient $\alpha \cong 1$. This does not match up with data analysis of astrophysics, where we have $\alpha \cong .2$ Those who work on these models report resolution of the Chaplygin Gas model for later time evolution when $\alpha < 1$, but as mentioned in the first section, the Chaplygin Gas model predicts when dark energy - dark matter unification is achieved through an exotic background fluid whose equation of state is given by p = - A/ρ$^\alpha$, and with $0 < \alpha \leq 1$. This was investigated via equivalent conditions of the equation of state involving varying parameter values of $w(z) = \frac{P}{\rho}$ from zero to -1, with the -1 value corresponding to traditional models involving the 'Einstein cosmological constant'

Secondly, is that one has undefined scale factors $a(t)$ for times less than Planck's time, $t << t_P$.

Thirdly, is that Brane world models will not permit Akshenkar's quantum bounce. [6,7] The quantum bounce idea is used to indicate how one can reconcile axion physics with the production of dark matter/dark energy later on in the evolution of the inflationary era where one sees Guth style chaotic inflation for times $t \geq t_P$ and the emergence of dark energy during the inflation era.

In addition is the matter of Sean Carroll, J. Chens paper [2] which pre supposes a low entropy – low temperature pre inflationary state of matter prior to the big bang. How does one ramp up to the high energy values greater than temperatures $10^{12}$ Kelvin during nucleosynthesis?

Fifthly, is the issue of relic graviton production. This is in tandem with a model which indicates a resolution of the issue which was raised by Guth, in 2003 in UCSB [12] about the multitude of string theory vacuum states, and if or not one preferred vacuum state could exist at the nucleation of a universe.

We need to investigate if or not gravity wave/ graviton generating functionals are congruent with Abbay Akshenkar's supposition for a quantum bounce. If they are, it would lead credence to Akshenkar's supposition [6, 7] of an earlier universe imploding due to contraction to the point of expansion used in measuring the birth of our universe.

Gravitons would appear to be produced in great number in the $\Delta t \approx t_P$ neighborhood, according to a brane world interpretation just given. This depends upon the temperature dependence of the 'cosmological constant.' And is for a critical temperature $T_C$ defined in the neighborhood of an initial grid of time $\Delta t \approx t_P$. Here, $T \equiv T_c \sim 250 \; GeV \Rightarrow N(T_c) \cong 51.5$. This among other things leads to a change in volume along the lines of, to crude first approximation imputing in numerical values to obtain

$$V = volume = \frac{5.625 \times 10^{57}}{T^6} \cdot \frac{1}{N^{3/2}(T)} \qquad (138)$$

The radius of this 'volume' is directly proportional to $3 \cdot t$ (setting the speed of light c =1). Note that we are interested in times t < $\Delta t \approx t_P$ for our graviton production, whereas we have a phase transformation which would provide structure for Guth's quadratic powered inflation.

A Randall-Sundrum effective potential, as outlined herein, would give a structure for embedding an earlier than axion potential structure, which would be a primary candidate for an initial configuration of dark energy .This structure would, by baryogenesis, be a shift to dark energy. We need to get JDEM space observations configured to determine if WIMPS are in any way tied into the supposed dark energy released after a $\Delta t \approx t_P$ time interval.

In doing this, we should note the following. We have reference multiple reasons for an initial burst of graviton activity, i.e. if we wish to answer Freeman Dyson's question about the existence of gravitons in a relic graviton stand point.

Now for suggestions as to future research. We are in this situation making reference to solving the cosmological "constant" problem without using G. Gurzadyan and She-Sheng Xue's [37]approach which is fixed upon the scale factor $a(t)$ for a present value of the cosmological constant. We wish to obtain, via Parks method of linking four and five dimensional cosmological constants a way to obtain a temperature based initial set of conditions for this parameter, which would eliminate the need for the scale factor being appealed to, all together. In doing so we also will attempt to either confirm or falsify via either observations from CMB based systems, or direct neutrino physics counting of relic graviton production the exotic suggestions given by Wald [1] for pre inflation physics and/or shed light as to the feasibility of some of the mathematical suggestions given for setting the cosmological constant parameter given by other researchers. Among other things such an investigation would also build upon earlier works initiated by Kolb, and other scientists who investigated the cosmological 'constant ' problem and general scalar reconstruction physics for early universe models at FNAL during the 1990s

Doing all of this will enable us, once we understand early universe conditions to add more substance to the suggestions by Bonnor, as of 1997 [38] for gravity based propulsion systems.  As well as permit de facto engineering work pertinent to power source engineering for this concept to become a space craft technology.

Further work needs to be done to be done along the lines suggested by Bo Feng's work on CPT violation physics and Quintessence scalar fields for the evolution/ production of dark matter/ dark energy . As well as to try to find experimental verification of a preferred initial vacuum state for cosmic nucleation in the time regime for times $t \leq t_P$. In addition, we need to establish more phenomenology links to the possibility of relic graviton production in the history of the early universe which may indicate a necessary proof of a preferred vacuum initial state for cosmic nucleation physics. This would avoid the problems shown up in Chapyron gas models between theory and experimental mesurements. [39] In addition, it would also give a more through development of Axion models toward their contribution to dark energy in a relic form [40]

The conclusion of the oscillatory behavior of a Quintessence scalar field as related to in Eqn 137 is very crude, but it is the start of understanding how dark energy, as related to by axion mass could interact with normal baryonic matter, and should be sharply upgraded. We did the simplest analysis of this problem possible, and it needs computer simulations to make it more tractable, and closer to astrophysical simulations.